\documentclass{mem}
\usepackage{natbib}
\usepackage{txfonts}
\usepackage{balance}
\usepackage{graphicx}
\usepackage{flushend}
\usepackage[a4paper,breaklinks,pdftex]{hyperref}
\idline{75}{282}
\begin{document}
\def\teff{$T\rm_{eff }$}
\def\kms{$\mathrm {km s}^{-1}$}

\title{
Imprint of the galactic acceleration scale on globular cluster systems: Galaxies in the Fornax Cluster
}

   \subtitle{}

\author{Michal B\'ilek\inst{1,2,3}
   \and Michael Hilker\inst{3}
   \and Florent Renaud\inst{4}
   \and Tom Richtler\inst{5}
   \and Avinash Chaturvedi\inst{3}
   \and Srdjan Samurovi\'c\inst{6}
          }

\institute{
LERMA, Observatoire de Paris, CNRS, PSL Univ., Sorbonne Univ., 75014 Paris, France\\
    \email{michal.bilek@obspm.fr}
        \and
        Coll\`ege de France, 11 place Marcelin Berthelot, 75005 Paris, France
        \and
        European Southern Observatory, Karl-Schwarzschild-Strasse 2, 85748 Garching bei M\"unchen, Germany
        \and
        Department of Astronomy and Theoretical Physics, Lund Observatory, Box 43, SE-221~00 Lund, Sweden
        \and
        Departamento de Astronomia, Universidad de Concepci\'on, Concepci\'on, Chile
        \and
        Astronomical Observatory of Belgrade, Volgina 7, 11060 Belgrade, Serbia
}

\authorrunning{B\'ilek et al. }

\titlerunning{Breaks in GC systems of Fornax Cluster galaxies}

\date{Received: Day Month Year; Accepted: Day Month Year}

\abstract{Dark matter is required in galaxies at galactocentric radii that are larger than the $a_0$-radius, which is where the gravitational acceleration generated by baryons of the galaxy equals the constant $a_0=1.2\times 10^{-10}$\,m\,s$^{-2}$ known as the galactic acceleration scale. It was found previously for massive early-type galaxies that the radial number-density profiles of their globular cluster (GC) systems follow broken power laws and the breaks occur at the $a_0$-radii. We have newly analyzed the distribution of GCs around galaxies in the Fornax cluster in existing photometric catalogs. We found that 1) the coincidence between $a_0$-radii and the break radii of globular cluster systems is valid for early-type galaxies of all masses and, 2) this also applies to the red and blue sub-populations of GCs separately.

}
\maketitle{}

\begin{figure*}
\resizebox{\hsize}{!}{\includegraphics[clip=true]{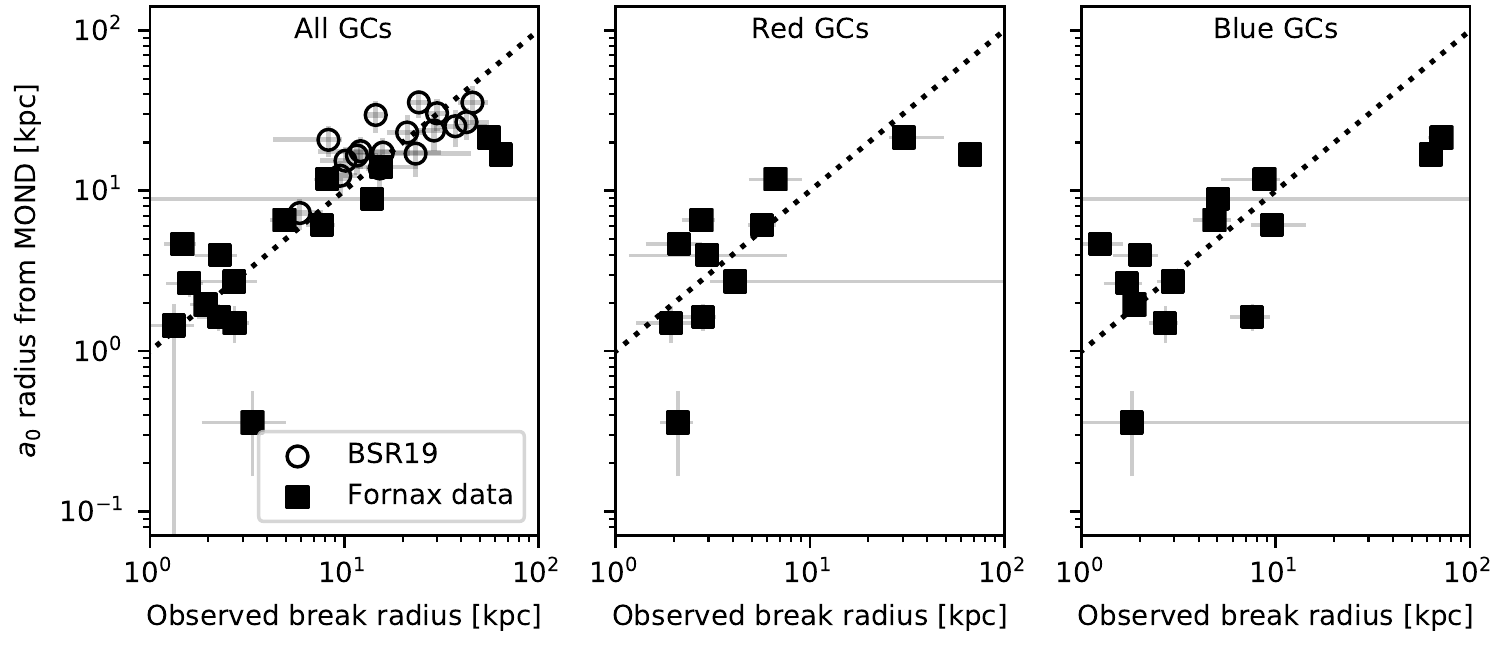}}
\caption{\footnotesize
Correlations of the $a_0$ and break radii. The black dashed lines indicate the one-to-one relation.
The errorbars indicate the 1-$\sigma$ uncertainty limits.}
\label{radii}
\end{figure*}


All galaxies, with the exception of the smallest dwarfs, are orbited by globular clusters (GCs, see, e.g., \citealp{brodie06} for a review). The GCs are compact ($\sim 3$\,pc) massive ($10^4-10^7\,M_\odot$) spherically symmetric star clusters. Their ages typically exceed 10\,Gyr, but GCs keep forming until today. The colors of old GCs follow a bimodal distribution. We thus divide GCs in the red (metal-rich) and blue 
(metal-poor) subpopulations, since they probably have different origins. The collection of all GCs that a galaxy has is called a GC system (GCS).

Gravitational fields of galaxies show an interesting regularity: assuming Newtonian gravity, we need dark matter only in the regions of galaxies where the strength of gravitational field generated by the observable matter $a_\mathrm{N}$ equals the constant $a_0 = 1.2\times 10^{-10}$\,m\,s$^{-2}$ and in these  regions, the gravitational acceleration follows $\sqrt{a_\mathrm{N} a_0}$. This behavior has been confirmed for a wide variety of galaxies through different methods \citep[e.g.,][]{lelli17}, and was originally predicted theoretically by the MOND theory of modified gravity and/or inertia \citep{milg83a}. The constant $a_0$ is called the galactic acceleration scale. We call the radius at which the gravitational acceleration generated by baryons equals $a_0$ the $a_0$ radius.

As we detail below, we found in \citet{bil19b} (BSR19 hereafter) that the density profiles of GCSs of massive early-type galaxies follow broken power laws, while the breaks occur at the $a_0$ radii. Later, we verified this for a larger variety of galaxies and blue and red GC subpopulations.

We came to our finding by fitting the radial profiles of the number-density of GCSs by a broken power law in the form:

\begin{equation}
\begin{tabular}{ lcl } 
 $\rho(r) = \rho_0\ r^a$ & for & $r<r_\mathrm{br}$, \\ 
$\rho(r) = \rho_0\ r_\mathrm{br}^{a-b}\ r^b$ &  for & $r\geq r_\mathrm{br}$.  
 \label{eq:bpl}
\end{tabular}
\end{equation}

The reason why we chose this profile stems from \citet{bil19}, where it was chosen because it allowed simple dynamical modeling of the GCSs and it described the data well. 
We call the parameter $r_\mathrm{br}$ at which the profile changes the slope as the break radius. Its physical meaning is not clear yet.

So far, we have analyzed the profiles of  about 40 GCS of elliptical and lenticular galaxies. The first sample of 17 massive early-type galaxies (ETGs) with spectroscopically confirmed GCs was presented in BSR19. The stellar galaxy masses ranged almost only between $10^{10}-10^{11}\,M_\odot$. The sample was a compilation of galaxies with a large number of spectroscopic GCs coming from different sources, often from the SLUGGS survey \citep{Brodie2014}.  In a submitted work (B\'ilek et al.), we used photometric catalogs of GC candidates for galaxies in the Fornax cluster, coming from ground-based Fornax Deep Survey  (FDS) \citep{cantiello20} and HST observations \citep{jordan07}. We used all ETGs covered by the FDS whose stellar masses exceed $10^{8}\,M_\odot$ (like the Magellanic clouds). The most massive galaxies had stellar masses of $10^{11}\,M_\odot$. The GCSs of dwarf galaxies of similar masses were stacked. We divided the GCs into blue and red according to their $g-z$ color \citep{fahrion20}.

 We derived a (complex) expression that gives the \textit{surface} number-density profile associated to Eq.~\ref{eq:bpl}. We divided the area around each investigated galaxy in annuli containing a fixed number of GCs. The observed surface number-density of GCs was then fitted by the analytic expression for the surface number-density profile using the maximum likelihood method. The uncertainties of the fit parameters were derived using the method of support.

For each galaxy, we considered $a_0$ radii calculated in two different ways: first, by using the Newtonian prescription for the gravitational field and second, by using the MOND prescription for the gravitational field, even though the two types of $a_0$ radii come out rather similar.
We found that the break radii occur near the $a_0$ radii (calculated in either way), see Fig.~\ref{radii}. The break and $a_0$ radii agree within the factor of two (the intrinsic scatter). This holds true for the total population of GCs and the red and blue GC subpopulations and this does not depend on whether we consider the $a_0$ radius calculated from the Newtonian or MOND gravity. The ratio of the break and $a_0$ radii is closer to unity for the MOND $a_0$ radii. 

The difference between the break radii of the red and blue GCs is not statistically significant.  We found this by fitting the distribution of this difference by a Gaussian distribution. The average is $-0.3^{+0.2}_{-0.2}$\,kpc, meaning that the break radius is insignificantly  higher for the blue subpopulation. The intrinsic scatter of the distribution came out $0.03^{+0.3}_{-0.02}$\,kpc. This result is based on a sample of 11 galaxies for which the break radii were available for both GC subpopulations.

We considered many  possible interpretations of the match between the break and $a_0$ radii, such as the consequence of galaxy mergers, of the MOND external field effect \citep{bm84}, dynamical friction, and others. Some of them have been proposed already in BSR19. In the submitted work, these and additional interpretations were explored in more detail. It turned out that none of them is perfect. For example, they do not induce broken power-law number-density profiles or cannot account for the fact that even dwarf galaxies, that are not expected to have experienced many mergers, have GCS number density profiles that follow broken power laws.

\begin{acknowledgements}
MB acknowledges the support by the ESO SSDF grant 21/10.  FR acknowledges support from the Knut and Alice Wallenberg Foundation. SS acknowledges the financial support of the Ministry of Education, Science and Technological Development of the Republic of Serbia through the contract No.~451-03-68/2022-14/200002.
\end{acknowledgements}

\bibliographystyle{aa}
\bibliography{citace}

\end{document}